
%
%
\def\ctc{closed timelike curve}
\def\ctcs{closed timelike curves}
\def\enrgy{energy momentum tensor}
\def\vev{vacuum expectation value of the energy momentum tensor}
\def\ph{polarised hypersurface}
\def\phs{polarised hypersurfaces}
\def\CH{Chronology horizon}
\font\title=cmr12
\magnification=\magstep1 \openup 2\jot
\rightline{DAMTP-R92/35}
\vskip-0.06in
\rightline{September 1992}
\vskip.2in
\centerline{\title Cosmic Strings and Chronology Protection}
\vskip.2in
\centerline{James D.E. Grant$^{\dagger}$}
\vskip.15in
\centerline{Department of Applied Mathematics and Theoretical Physics,}
\vskip-0.06in
\centerline{University of Cambridge, Silver Street, Cambridge CB3 9EW, United
Kingdom.}
\vskip 0.1in
\centerline{Theoretical Astrophysics, Californian Institute of Technology,}
\vskip-0.06in
\centerline{Pasadena, California 91125.}
\vskip0.3in
\centerline{Abstract}
\vskip0.2in
\noindent A space consisting of two rapidly moving cosmic strings has recently
been constructed by Gott that contains closed timelike curves. The global
structure of this space is analysed and is found that, away from the
strings, the space is identical to a generalised Misner space. The vacuum
expectation value of the energy momentum tensor for a conformally coupled
scalar field is calculated on this generalised Misner space. It is found to
diverge very weakly on the {\CH}, but more strongly on the {\phs}. The
divergence on the {\phs} is strong enough that when the proper geodesic
interval around any {\ph} is of order the Planck length squared, the
perturbation to the metric caused by the backreaction will be of order one.
Thus we expect the structure of the space will be radically altered by the
backreaction before quantum gravitational effects become important. This
suggests that Hawking's `Chronology Protection Conjecture' holds for spaces
with non-compactly generated {\CH}.
\vfill
\eject

\beginsection{1. Introduction}

It has long been known that changes of spatial topology give rise to either
singularities or {\ctcs} [1]. It was therefore not too surprising, with
hindsight, when several topologically non-trivial spaces were recently
constructed that generically contained {\ctcs} [2-4]. These spaces were
formed by removing two spheres from a spacetime and then  joining the
resulting holes together by a cylinder to form a ``wormhole''. It was found
thatif one of the wormhole mouths was in a generic gravitational field, or if
the wormhole mouths were in generic relative motion, then {\ctcs} would form.
One of the major drawbacks of these spaces, however, is that they are
not vacuum solutions of Einstein's equations, and the matter required to
maintain the spaces must violate an averaged form of the Weak Energy Condition
[3].

A simpler space with {\ctcs} has now been constructed by Gott [5]. Gott's
space just contains two cosmic strings moving past each other at high
velocity. The space is locally flat away from the strings, so there is no
violation of the Weak Energy Condition, and the topology is just $R^4$.

The aim of the present paper is to study the quantum mechanical stability of
the Gott spacetime. We consider putting a conformally coupled scalar field
into the Gott space background and calculate the vacuum expectation value of
its {\enrgy}. Early calculations of such quantities in spaces with {\ctcs}
[6] suggested that such quantities would diverge at the {\CH}
(the boundary of the region containing {\ctcs}). It was then
shown in [7] that in any spacetime with {\ctcs}, the {\CH} is
the limiting surface of a family of polarised hypersurfaces, and that the
{\enrgy} of the field will diverge on all of the {\phs} (See
also [8] for a specific example). This means that in any space with {\ctcs}
there will be surfaces arbitrarily close to the {\CH} where the {\enrgy} is
divergent. If this divergent {\enrgy} is now used as a source term in the
semi-classical Einstein  equations,  $R_{ab} - {1\over 2} R g_{ab} = 8\, \pi\,
l_P^2\langle T_{ab}\rangle$, then we expect the backreaction to radically alter
the spacetime round about the {\CH}, and stop us reaching the causality
violating region.

In Gott's space, we find that the divergence of the {\enrgy} is very weak at
the {\CH}. The perturbation of the metric due to the backreaction of this
divergence would be unobservable even when we are a Planck length, $l_P$,
from the {\CH}. However, the divergence is stronger as we approach the
{\phs}. Here we find that when the proper geodesic distance squared around any
polarised hypersurface, ${\sigma}_n$ for some integer $n$, is of order $l_P^2$,
then the metric perturbation is of order $1$. This will radically alter the
structure of the spacetime, and suggests that Hawking's `Chronology Protection
Conjecture', originally only meant to apply to spaces with compactly generated
{\CH}s, will also apply in the non-compactly generated case.

We begin in section 2 with a brief review of the Gott construction, and
section 3 is devoted to an analysis of the space's geometrical properties.
The conclusion here is that, far from the strings, the space is identical to
a generalised Misner universe, the relevant properties of which are then
reviewed. We then calculate, in section 4, the {\vev} for a conformally
coupled scalar field on this generalised Misner space space. In section 5 we
discuss these results and their implications for Hawking's `Chronology
Protection Conjecture'.
%
%
%
%
\beginsection{2. Gott's Cosmic String Spacetime}

We consider a space containing an infinitely long, straight cosmic string.
This can be viewed as flat Minkowski space with a wedge of angle $2\alpha$
cut out along the axis of the string. We can choose Minkowski coordinates,
$(t, x, y, z)$, and place the core of the string on the line $x=0, y=d$,
with $z$ as the coordinate along the axis of the string. We remove the
wedge from the space so that points with $x = \pm\ (y-d)\ \tan \alpha$
are identified (see fig.1).

Suppose we now consider two points $A$ and $B$ at rest on the surface $y=0$,
where $x_{A}^{\ a} = (t,x_0,0,0),\ x_{B}^{\ a} = (t,-x_{0},0,0)$. There are
now two paths that a light signal sent from $A$ could follow to arrive at
$B$. The first would be the direct path $AOB$. If $x_0$ is big enough,
then there is an alternative path, $ACDB$, that goes around the cosmic
string, making use of the angular deficit. If $x_0 \tan \alpha >> d$, then
this second light ray will arrive at $B$ before the direct one.

If a light beam travelling around the cosmic string can arrive before
the light beam passing through $O$, then so can a rocket travelling at
sufficiently high velocity. The event of the rocket leaving $A$,
$x_i$, and arriving at $B$, $x_f$, will be spacelike separated, since the
light ray travelling along $y=0$ arrives at $B$ after event $x_f$. Hence
we can find a Lorentz frame, in which the string moves at velocity $v$
in the $+x$ direction, in which the events $x_i$ and $x_f$ are simultaneous,
i.e. the rocket is seen to arrive at point $B$ at the same time as it left
point $A$.

We can take two copies of the above space and glue them together along their
$y=0$ surfaces. We boost the region $y \ge 0$ at velocity $v$ in the $+x$
direction, and the region $y\le 0$ at velocity $v$ in the $-x$ direction.
Physically, if we are in the centre of mass frame, all this means is that we
see two cosmic strings going in opposite directions at speed $v$, with impact
parameter $2d$ (see fig. 2).

The construction above showed that in the centre of mass frame, we could
see a rocket leaving event $x_i$ and simultaneously arriving at event
$x_f$ if it followed the path $ACDB$. If the rocket then turns around
at $x_f$, then by the same argument it can travel back round the lower
cosmic string by path $BEFA$ and arrive back at event $x_i$. We have thus
created a {\ctc} through event~$x_i$.

There is, however, a restriction on the velocity $v$. It can be shown [5]
that, if $x_0>>d$, then we need

$$ \cosh \xi \sin \alpha > 1 \eqno(2.1) $$

in order to get \ctcs, where $v=\tanh \xi$. Grand Unified Theories
usually predict $\alpha \approx 10^{-5}$, meaning that $v \approx
c\,(1-10^{-10})$. This may seem rather unrealistic, but it is possible that
cosmic strings created in the early universe would have such high velocities.
%
%
%
%
\beginsection{3. Geometry Of The Space}

Following [9], we now look for a more geometrically transparent
representation of the above construction.

If one considers parallel transport of vectors around a closed curve in
a spacetime that includes a cosmic string, then there is a non trivial
holonomy if the closed curve encloses the string. If the string is at rest,
this holonomy is just a rotation through angle $2\alpha$, where $2\alpha$
is the deficit angle of the string. If the string is moving at constant
velocity $v$ in the positive $x$ direction, then the holonomy is
represented by the matrix $B(v)\ R(2\alpha)\ B(-v)$, where $B(v)$ is
the boost matrix corresponding to velocity $v$, and $R(2\alpha)$ is
the matrix corresponding to a rotation through angle $2\alpha$. In the
case of the Gott spacetime, the holonomy matrix for a closed curve
around both strings will be

$$ H(v,\alpha) = (B(-v)\, R(2\alpha)\, B(v))\, (B(v)\, R(2\alpha)\, B(-v)).
\eqno(3.1) $$

We would like to know if this corresponds to a rotation, or a boost.
Therefore we consider the trace of $H$, which will be less than
$4$ if $H$ corresponds to a rotation and greater than $4$ if $H$ corresponds
to a boost. If we take $\tanh \xi = v$, then we find

$$ Tr(H) - 4 = 8 \cosh^2 \xi\ (1 - \cos 2\alpha)
\ (\sin^2 \alpha \cosh^2 \xi - 1 ).\eqno(3.2)$$

The Gott space has {\ctcs} if $\sin \alpha \cosh \xi > 1$, and thus
corresponds to a holonomy of a boost.

The Gott space is flat away from the strings. The holonomy around a closed
curve that encloses both cosmic strings is a boost. This suggests that one
could view the region with {\ctcs} as flat Minkowski space identified under
the action of a boost.

The exact identification we require can be found by tracing a curve in the
Gott space that would usually close up in flat space. This is similar to
the case with one cosmic string where a curve that would close up in flat
Minkowski space will not close up if the curve goes around the string. The
amount the curve does not close up by is a rotation about the axis of the
string, through the deficit angle of the string. Therefore in this case, the
amount that the curve does not close up by is the same as the holonomy
around the string. In general the endpoint of the curve will be
$x' = Hx + C$, where $H$ is the holonomy around any closed curve
enclosing the string, and $C$ is some constant vector.

Defining the function

$$ \Delta = \cosh^2\xi \sin^2 \alpha - 1,\eqno(3.3)$$

and expressing the results in terms of the coordinates

$$ t' = {1 \over{\Delta }}\,[ t \sinh \xi\, \sin \alpha - y \cos \alpha]
- {d\over 2}\cosh^2 \xi\,\sin 2\alpha \,(4\,\sin^2 \alpha + (2\Delta + 1)^2),
\eqno(3.4)$$

$$ x' = x\, -\, {d\over{\Delta}}\cosh \xi\, \sin \alpha\, \sin
2\alpha\, (2\Delta + 1)\,(1 + \cosh^2 \xi \,\Delta),
\eqno(3.5)$$

$$ y' = {1 \over {\Delta }}\,[ y \sinh \xi\, \sin \alpha - t \cos \alpha],
\eqno(3.6)$$

\noindent it is straightforward to show that the components of $H$ and $C$
are given by

$$
H (\alpha,\xi) = \left(\matrix
{ 1+f(\alpha,\xi)\Delta&g(\alpha,\xi)\Delta&\ \ \ 0\ \ \ &\ \ \ 0\ \ \ \cr
                     g(\alpha,\xi)&1+f(\alpha,\xi)\Delta&0&0\cr
                     0&0&1&0\cr
                     0&0&0&1\cr}\right),
\eqno(3.7)$$

$$ C^{y'} =\,-\,{4d\over{\Delta}}\,\sinh \xi\,\sin \alpha,\eqno(3.8)$$

$$ C^{t'} = C^{x'} = C^{z} = 0.\eqno(3.9)$$

\noindent Here we have defined the functions

$$ f(\alpha,\xi)\,=\,4\,\cosh^2 \xi \,(1\,-\,\cos 2\alpha),\eqno(3.10)$$

$$ g(\alpha, \xi)\,=\,4\,\cosh \xi\,\sin \alpha \,
(\cosh^2 \xi \,(1\,-\,\cos 2 \alpha)\,-\,1).\eqno(3.11)$$

If we assume there are \ctcs, then (2.1) implies that $\Delta > 0$. If we
change to coordinates $\tilde t = \Delta^{1/2}\ t',\ \tilde y =
\Delta^{1/2}\ y'$, the metric becomes the flat space metric $ds^2 = -d \tilde
t^2 + dx'^2 + d \tilde y^2 + dz^2$. In terms of these
coordinates, the holonomy takes the form of a boost in the $\tilde t - x'$
plane with parameter $a$ given by

$$ \cosh a\,=\,1\,+\,f(\alpha,\xi)\,\Delta,\eqno(3.12)$$

\noindent and the vector $C$ becomes a displacement in the
$\tilde y$ direction of distance $b$, where

$$ b\,=\,-\,{4d\over{\Delta^{1/2}}}\,\sinh \xi \,\sin \alpha.\eqno(3.13)$$

Thus, for any observer that travels around both strings, the Gott space
will  be physically indistinguishable from flat space identified under
the combined discrete action of a boost in the $\tilde t-x'$ plane and a
translation in the $\tilde y$ plane. (From now on we drop the tildes and
primes on these coordinates.)

This is just a generalisation of Misner space [10,11]. Here we pick an
origin, $O$, in flat 2 dimensional Minkowski space, and identify the points
$A^n(x)$, for all integers $n$ and $x\in J^{-}(O)$, where

$$ A^n(x)\equiv (t \cosh na +x \sinh na ,\, x \cosh na  + t \sinh na ).
\eqno(3.14)$$

Under a Lorentz boost of velocity $v = \tanh a$, the point $x$ is carried to
the point $A(x)$. Thus, physically, Misner space corresponds to the bottom
quadrant of Minkowski space identified under the action of a discrete boost.

Introducing coordinates $T$ and $X$ such that $t=-T\cosh X,\ x=-T\sinh X$,
the metric becomes $ds^2 = - dT^2 + T^2 dX^2$ and the above identified
region corresponds to $T>0$ with coordinate $X$ having period $a$. We can
extend this metric through the surface $T=0$, where it becomes degenerate,
by introducing coordinates $\tau = T^2,\ u = X - \log T$, giving the metric
$ds^2 = du\  d\tau + \tau\ du^2$, which is non-degenerate for all
real $\tau$. The region $\tau < 0$ now contains \ctcs, and the surface
$\tau = 0$ contains closed null geodesics. This extended space corresponds
to the bottom and left hand wedges of Minkowski space identified under the
action of the boost defined by $A$ above. One can do similar extensions and
consider the whole of the two dimensional Minkowski space being identified
under this discrete boost. In order for the resulting space to be a manifold,
however, we must delete the origin, $O$. The resulting manifold is then
non-Hausdorff and the space is geodesically incomplete [11].

Two flat dimensions can now be added to the above space. We then have the
freedom to make identifications in these extra directions. The discussion
above suggests that the Gott space is the same as four dimensional Minkowski
space with the points $B^n(x)$ identified, for all integers $n$, where

$$ B^n(x) \equiv
(t \cosh na +x \sinh na ,\, x \cosh na  + t \sinh na ,\, y + nb,\, z),
\eqno(3.15)$$

\noindent and $a$ and $b$ are defined by (3.12) and (3.13) respectively.
As long as $b \neq 0$, no points need be removed from the space. The
resulting manifold is Hausdorff, and tghe space is geodesically complete.
The fact that the Gott space has {\ctcs} at any value of $t$ [12] is now
analagous to the fact that the identified left (and right) hand quadrant of
identified Minkowski space have {\ctcs} at arbitrary values of the Minkowski
coordinate $t$. Further, the fact that in Misner space the surfaces
$\tau = {\rm constant} > 0$ are not intersected by any {\ctcs} suggests that
there will exist similar achronal surfaces in the Gott space which are not on
any {\ctcs} [13].

Any point in the covering Minkowski space is null separated from another
copy of the same point if its coordinates satisfy

$$ x^2 - t^2 = {n^2\,b^2 \over {2\ ( \cosh na - 1)}}.\eqno(3.16) $$

\noindent Thus every point on this surface can be joined to itself by a
(unique) null geodesic that passes around both strings $n$ times. Although,
in the physical space, the above null geodesic passes through the same
point twice, its tangent vector differs on these two occasions by the
holonomy $H^n$. Following [13] and [7] respectively, we call such lines
{\it self intersecting null geodesics} and call the surface defined by
(3.16) the {\it $n$'th  polarised hypersurface}. We reserve the term
{\it closed null geodesic} for a line whose tangent vector coincides
at the point each time, and thus goes through the point an infinite
number of times.

If we take the limit $n \to \infty$ in (3.16) then we find that the
chronology horizon is situated at $t = \pm x$. This is a null surface,
but unlike in the ordinary Misner space case it does not contain any
closed null geodesics. This is because any null geodesic in the surface
must have $y={\rm constant}$, and  so cannot join two identified points.
Therefore the null geodesics that  generate the horizon will never enter
and remain within a compact region when followed backwards in time. So,
in the terminology of [14] the {\CH} is non-compactly generated.
%
%
%
%
\beginsection{4. Matter Fields On The Space}

\indent We here use the results of the preceding sections to consider
placing quantum mechanical matter into the Gott space. For simplicity we
take a conformally coupled scalar field, and we calculate the {\vev}
for this field, $\langle T_{ab}\rangle$. In a flat four dimensional
space, the renormalised propagator of a scalar field is

$$ G(x,x') = {1\over {(2 \pi)^2}}
\sum_{n=-\infty \atop n \neq 0}^{+\infty}\sigma_n (x,x')^{-1},\eqno(4.1)$$

\noindent where $\sigma_n (x,x')$ is the square of the proper geodesic
distance from  $x$ to $x'$ along the $n$'th geodesic joining the two points.
To calculate $\langle T_{ab}\rangle$, we differentiate this propagator twice
with respect to position and take the limit $x'\to x$ (see (4.5)). Thus we
would expect $\langle T_{ab}(x)\rangle$ to behave like

$$\langle T_{ab}(x)\rangle \sim \lim_{x' \to x}\,
\sum_{n=-\infty \atop n \neq 0}^{ \atop +\infty}\, \sigma_n (x,x')^{-3}.
\eqno(4.2)$$

If there is a self intersecting null geodesic through $x$, then one of the
$\sigma_n (x,x)$ will vanish, and so $\langle T_{ab}\rangle$ will diverge at
$x$. If one now makes a semi-classical approximation and treats $\langle
T_{ab}\rangle$ as a source term in the Einstein field equations, one might
hope that the divergence of $\langle T_{ab}\rangle$ on the {\phs} would
induce a singularity, making these surfaces non-traversable. Since there are
{\phs} arbitrarily close to the {\CH}, we would therefore hope that these
divergences would also make the {\CH} non-traversable. This is the basis
upon which Hawking put forward the `Chronology Protection Conjecture' which
states that {\ctcs} cannot be created [14].

In the covering space of Gott space, we begin with the ordinary flat space
propagator

$$ G_{0}(x,x') = {1\over {(2 \pi)^2}} [\
-(t-t')^2+(x-x')^2+(y-y')^2+(z-z')^2\ ]^{-1}.\eqno(4.3)$$

The renormalised propagator on the identified spacetime is then

$$
\eqalignno{G(x,x') = {1\over {(2 \pi)^2}}
\sum_{n=-\infty \atop n \neq 0}^{+\infty}
[\ -(t-(t'\cosh na + x'\sinh na))^2 &+(x-(x'\cosh na + t'\sinh na))^2\cr
+(y-(y'+nb))^2&+(z-z')^2\ ]^{-1}.&(4.4)\cr}$$

This propagator is already symmetric under interchange of $x$ and $x'$, so
we obtain the renormalised {\enrgy} of the field [15] from

$$ \langle T_{ab}\rangle =\lim_{x' \to x} [\, {2 \over 3}\, \nabla_a
\nabla_{b'} -  {1 \over 3}\, \nabla_a \nabla_b - {1 \over 6}\, g_{ab}
\nabla_{c'}\nabla^{c'}]\, G(x,x').\eqno(4.5)$$

On carrying out this calculation with the above propagator, one finds that
the only non-zero components of $\langle T^{a}_{\ \ b}\rangle$ are

$$ \langle T^{T}_{\ T}\rangle = {1\over{3\pi^2}} \sum_{n=1}^{\infty}\
{( \cosh na + 2)\over {f_n^2}},\eqno(4.6a)$$

$$ \langle T^{X}_{\ X}\rangle = {1\over{3\pi^2}} \sum_{n=1}^{\infty}\
{(\cosh na + 2)\over {f_n^2}}\ \Bigl[\ -3\ +\ {4 n^2 b^2 \over
{f_n}}\ \Bigr],\eqno(4.6b)$$

$$ \langle T^{y}_{\ y}\rangle = {1\over{3\pi^2}}
\sum_{n=1}^{\infty}\ \Bigl[\ {( \cosh na + 2)\over {f_n^2}}\
-\ {2n^2 b^2 (\cosh na + 5)\over {f_n^3}}\ \Bigr],\eqno(4.6c)$$

$$ \langle T^{z}_{\ z}\rangle =
{1\over{3\pi^2}} \sum_{n=1}^{\infty}\ \Bigl[\ {( \cosh na + 2)\over {f_n^2}}\
-\ {2 n^2 b^2 (\cosh na - 1)\over {f_n^3}}\ \Bigr],\eqno(4.6d)$$

\noindent where $t = T \cosh X,\ x = T \sinh X$ and

$$ f_n = 2\, (t^2 -
x^2)\, (\cosh na - 1) + n^2 b^2 = 2\, T^2\, (\cosh na - 1) + n^2 b^2.
\eqno(4.7)$$

These expressions diverge on the {\CH}, where $T=0$, and on the
{\phs}, where $f_n = 0$ for some integer $n$. (We note in passing that the
$(T,X,y,z)$ coordinate system becomes singular at $T=0$, the {\CH}.)

If we approach the {\CH}, we can approximate the above sums by
integrals, and evaluate the asymptotic behaviour by a saddle point method.
We find that the components of the {\enrgy} diverge like $K / b^2 T^2$,
where $K$ is a negative constant. We can estimate the perturbation this will
cause in the metric by using it as a source term in the semi-classical
Einstein equations, %
$R_{ab} - {1 \over 2}\, R\, g_{ab} =
8\, \pi\, l_P^2\, \langle T_{ab}\rangle$, %
where $l_P$ is the Planck length. Therefore the perturbation to the
curvature will be of order $Kl_P^2 / b^2 T^2$ [16]. To find the metric
perturbation felt by someone travelling along a geodesic  $(X, y,
z)\,=\,constant$, we have to integrate twice with respect to $T$, giving

$$ \delta g \approx {K\,l_p^2\over {b^2}}\,\log(T/b),\eqno(4.8)$$

This perturbation diverges at the {\CH}. However, we expect quantum
gravity  will come into play before the horizon, and may smooth out the
divergence  before it becomes noticeable. It seems reasonable to assume
that quantum gravitational effects will become important at some Lorentz
invariant,  observer independent distance from the horizon [14]. Thus a
first  approximation would be that quantum gravitational effects come
into play when

$$ T \approx l_P.\eqno(4.9) $$

Putting this into (4.8), and assuming that $b$ is some typical
macroscopic distance of order one metre, gives a metric perturbation
of order $10^{-70}$. This would be completely unobservable.

We can study the behaviour of $\langle T^{a}_{\ \ b}\rangle$ near the
{\phs} by defining a coordinate $\tilde T$ by $\tilde T^2=x^2-t^2$.
We find  that the components of $\langle T^{a}_{\ \ b}\rangle$ diverge,
at worst, like $K'\,b^2/(\tilde T+\tilde T_n)^3\,(\tilde T-\tilde T_n)^3$
as we approach the $n$'th \ph, where $\tilde T_n$ is the value of
$\tilde T$ on that surface, and $K'$ is a constant. This means that
the dominant contribution to the metric perturbation caused by the
back reaction of the matter will be

$$ \delta g_n \approx
{K'\, l_p^2\, b^2 \over {(\tilde T+\tilde T_n)^3\,(\tilde T-\tilde T_n)}},
\eqno(4.10)$$

It is more difficult to estimate when quantum gravitational effects
become  strong close to the the {\phs}. If we were to treat the
gravitational field  like a massless spin 2 field in flat spacetime,
we would expect the quantum fluctuations of the field to be governed
by the geodetic interval around the polarised hypersurface ${\sigma}_n(x,x)$
[17]. This suggests that quantum  gravity would become significant when
$(\tilde T+\tilde T_n)\,(\tilde T-\tilde T_n)\,\approx\,l_P^2$. This
leads to a metric perturbation of order $1$, which would radically
alter the structure of the space around the {\phs} and the
{\CH}\footnote{$^1$}{I am grateful to Kip Thorne for this argument;
see [18].}.

Thus, it appears that around the {\phs}, quantum gravity will not enter until
the metric perturbation has become large enough to change the structure of
the space.

\beginsection{5. Conclusion}

We have shown that, away from the strings, the Gott space is identical to
flat Minkowski space identified under the action of a discrete boost and
translation. On calculating the {\vev} for a conformally coupled scalar
field on this space, we find that it diverges on the {\CH} and on the {\phs}.
The divergence around the {\phs} is sufficiently strong that we expect the
backreaction of the field to radically alter the structure of the space
before quantum gravitational effects have come into play.

These results seem to extend Hawking's `Chronology Protection Conjecture'
which states that {\ctcs} cannot be created [14]. The Chronology Protection
Conjecture originally only refered to spaces where the region of {\ctcs} was
compact, but our results seem to suggest that it also applies in spaces with
non-compactly generated {\CH}s.

It remains to be shown that the Green function of the wave
equation given by (4.4) corresponds to the propagator of a physically
acceptable quantum state of the field on the  space (in the sense of [19]).
Indeed, there are serious problems in doing quantum field theory on any
non-globally hyperbolic spacetime (see [20] for one approach to this problem).
It has been shown, however, that the Green function constructed in [6] for
Misner space {\it is} a propagator for a real quantum state and that the
Hiscock-Konkowski state is actually a thermal state [21]. Hopefully, similar
arguments should apply to the present case.

\vskip 0.25in
\beginsection{Acknowledgements}

\indent I would like to thank Stephen Hawking and Kip Thorne for their
guidance and suggestions. I also wish to convey my thanks to the Theoretical
Astrophysics group in Caltech, who were so helpful and encouraging when this
work was being completed there. I also must thank St. John's College,
Cambridge, for a travel grant which enabled me to visit Caltech.

This research was carried out under a Science and Engineering Research
Council (SERC) studentship.

\bigskip
\hrule
\bigskip

\item {$^{\dagger}$\ }%
email address : jdeg1@phx.cam.ac.uk.
\item {[1]\ }%
R. P. Geroch, J. Math. Phys {\bf 8}, 782 (1967).
\item {[2]\ }%
M. S. Morris and K. S. Thorne, Am. J. Phys. {\bf 56}, 395 (1988).
\item {[3]\ }%
M. S. Morris, K. S. Thorne and U. Yurtsever, Phys. Rev. Lett.
{\bf 61}, 1446 (1988).
\item {[4]\ }%
V. P. Frolov and I. D. Novikov, Phys. Rev. {\bf D42}, 1057 (1990).
\item {[5]\ }%
J. R. Gott, Phys. Rev. Lett. {\bf 66}, 1126 (1991).
\item {[6]\ }%
W. Hiscock \& D. A. Konkowski, Phys. Rev. {\bf D26}, 1225 (1982).
\item {[7]\ }%
S.-W. Kim and K. S. Thorne, Phys. Rev. {\bf D43}, 3929 (1991).
\item {[8]\ }%
V. P. Frolov, Phys. Rev. {\bf D43}, 3878 (1991).
\item {[9]\ }%
S. Deser, R. Jackiw and G. t'Hooft, Annals of Physics, {\bf 152}, 220
(1984).
\item {[10]\ }%
C. W. Misner in {\it Relativity Theory and Astrophysics I:
Relativity and Cosmology}, edited by J. Ehlers, Lectures in
Applied Mathematics, Volume 8 (American Mathematical Society, 1967).
\item {[11]\ }%
{\it The Large Scale Structure of Space-time}, S. W. Hawking \&
G. F. R. Ellis (Cambridge University Press, Cambridge, 1973).
\item {[12]\ }%
A. Ori, Phys. Rev. {\bf D44}, 2214 (1992).
\item {[13]\ }%
C. Cutler, Phys. Rev. {\bf D45}, 487 (1992).
\item {[14]\ }%
S. W. Hawking, `The Chronology Protection Conjecture', DAMTP
preprint R91/15, submitted to Phys. Rev. D.
\item {[15]\ }%
R. M. Wald, Phys. Rev. {\bf D17}, 1477 (1978).
\item {[16]\ }%
Kip Thorne (private communication) has done similar calculations with the
{\enrgy} in [7] for the wormhole spacetime. Here the {\CH} is compactly
generated and thus contains a smooth closed null geodesic. He finds that
if the wormhole has no defocusing effect ($b/2D = 1$ in the terminology
of [7]), then $\langle T^{ab}\rangle$ diverges like the inverse square
of time $T$ to the horizon (note: this $T$ is analagous to the coordinate
$t$ in our terminology). Thus the divergence in this case is similar to
that in our generalised Misner space. However, if the defocusing is
non-zero ($b/2D \neq 1$) then the divergence is weaker and if $b/2D$ is
sufficiently small, there is no divergence on the {\CH} except on the closed
null geodesic. This may be the generic behaviour away from the closed null
geodesic on a space where the {\CH} is compactly generated.
\item {[17]\ }%
It may be helpful to say what we mean by $\sigma_n (x,x)$, when $x$ is
not on the $n$'th \ph. We can look on $x$ as the first of an infinite
set of points, $\{ x_i:i=1,2,3\dots\,:x_1=x\}$, that converge to some point
$y$ on the $n$'th \ph. Since $y$ is on the $n$'th \ph, there will be a  self
intersecting null geodesic, ${\cal G}$, through $y$. This self intersecting
null geodesic is the limit of a set of closed spacelike geodesics, ${\cal
G}_i$, through the corresponding points $x_i$. $\sigma_n (x,x)$ is then the
proper geodesic interval around the closed spacelike geodesic ${\cal G}_1$.
For this reason, we refer to $\sigma_n (x,x)$ as the proper geodesic interval
{\it around} the polarised hypersurface.  \item {[18]\ }%
K. S. Thorne in {\it Proceedings of 13th
International Conference on General Relativity and Gravitation}, ed.
C. Kozameh (Institute of Physics, Bristol, England, in press).
\item {[19]\ }%
B. S. Kay in {\it Differential Geometrical Methods in Theoretical Physics},
eds. K. Bleuler  and M. Werner (Kluwer Academic Publishers, 1988).
\item {[20]\ }%
B. S. Kay, `The Principle of Locality and Quantum Field Theory on
(Non-Globally Hyperbolic) Curved Spacetimes', DAMTP preprint R92/22, to be
published in Rev. Theor. Phys.
\item {[21]\ }%
B. S. Kay, unpublished.

\end